\documentclass{article}
\usepackage{spconf,amsmath,graphicx,hyperref}
\usepackage{booktabs}
\usepackage{multirow}
\usepackage{amssymb}
\usepackage{graphicx}  

\title{OMNI-CLST: ERROR-AWARE CURRICULUM LEARNING WITH GUIDED SELECTIVE CHAIN-OF-THOUGHT FOR AUDIO QUESTION ANSWERING}
\name{Jinghua Zhao\textsuperscript{1,2}, Hang Su\textsuperscript{2}, Lichun Fan\textsuperscript{2}, Zhenbo Luo\textsuperscript{2}, Hui Wang\textsuperscript{1}, Haoqin Sun\textsuperscript{1}, Yong Qin\textsuperscript{1}}
\address{$^{1}$TMCC, College of Computer Science, Nankai University, Tianjin, China \\ $^{2}$MiLM Plus, Xiaomi Inc., China}
\begin{document}
%
\maketitle
\begin{abstract}
With the rapid progress of large audio-language models (LALMs), audio question answering (AQA) has emerged as a challenging task requiring both fine-grained audio understanding and complex reasoning. While current methods mainly rely on constructing new datasets via captioning or reasoning traces, existing high-quality AQA data remains underutilized. To address this, we propose Omni-CLST, an error-aware \textbf{C}urriculum \textbf{L}earning framework with guided \textbf{S}elective Chain-of-\textbf{T}hought. The framework efficiently leverages existing high-quality dataset through two key strategies: an error-aware curriculum that organizes samples by difficulty, and a guided thought dropout mechanism that focuses reasoning on challenging cases. Experiments show that Omni-CLST achieves 73.80\% on MMAU-mini and a new state of the art of 64.30\% on MMAR, demonstrating robust generalization in multimodal audio-language understanding.
\end{abstract}
\begin{keywords}
Audio Question Answering, Curriculum Learning, Chain-of-Thought, Group Relative Policy Optimization
\end{keywords}
\section{Introduction}
\label{sec:intro}

Large Audio-Language Models (LALMs)~\cite{chu2024qwen2, xu2025qwen2, ding2025kimi, pmlr-v235-kong24a, ghosh2025audio, goel2025audio, google2025gemini} have developed rapidly in recent years, pushing the boundaries of multimodal audio understanding. Beyond Automatic Audio Captioning (AAC), a new task, Audio Question Answering (AQA) has emerged as a more challenging benchmark for comprehensive audio understanding and reasoning. It requires models to generate accurate answers by selecting from given options to natural language questions grounded in audio inputs.

Recent efforts have attempted to enhance AQA performance by incorporating deep reasoning into LALMs through Chain-of-Thought (CoT). 
Audio-CoT~\cite{ma2025audio} is the first to evaluate CoT reasoning in AQA based on Qwen2-Audio, highlighting the importance of transferring reasoning capabilities into multimodal audio understanding systems, though the improvements remained limited. 
Audio-Reasoner~\cite{xie2025audio} further constructs a structured CoT process using Gemini and diverse speech, sound, and music datasets, achieving significant gains, albeit with potentially redundant reasoning steps. 
Building on the reasoning advances of DeepSeek-R1~\cite{guo2025deepseek}, R1-AQA~\cite{li2025reinforcement} applies the GRPO (Group Relative Policy Optimization) algorithm to LALMs for the first time, substantially enhancing reasoning capabilities. 
Inspired by R1-AQA, subsequent works extend GRPO training to LALMs using different strategies.
SARI~\cite{wen2025sari} constructs structured CoT datasets to provide stronger reasoning supervision and employs a sophisticated curriculum learning strategy to further enhance model training.
AudSemThinker~\cite{wijngaard2025audsemthinker} extracts semantics of sound via multiple classifiers and incorporats them into CoT reasoning.
Omni-R1~\cite{rouditchenko2025omni} generates new QA pairs from captions with ChatGPT~\cite{achiam2023gpt} to enrich the training corpus.
The follow-up work of AudSemThinker~\cite{wijngaard2025data} attempts to employ a sophisticated curriculum learning strategy but does not yield significant overall gains.

However, these approaches face several challenges. First, constructing new datasets is costly and time-consuming, while existing high-quality AQA datasets remain underexploited, and the reliance on captioning models as the sole source of audio understanding leads to homogeneous QA pairs with limited diversity. Second, reinforcement learning with GRPO is computationally intensive and slow to converge, which makes it inefficient to use the full dataset. In practice, most works randomly subsample a portion of data for GRPO training, yet it remains unclear which types of data are more suitable to maximize training efficiency and model improvement. A more principled strategy for data selection could substantially enhance GRPO effectiveness. Third, although Chain-of-Thought (CoT) supervision has shown promise in improving reasoning abilities, it is still an open problem how to best utilize CoT-annotated data—naively applying CoT to all samples may even interfere with cases solvable without explicit reasoning.

Motivated by the above challenges, we propose Omni-CLST, an error-aware Curriculum Learning framework with guided Selective Chain-of-Thought. The framework is built upon Qwen2.5-Omni. Specifically, Omni-CLST first applies a filtering strategy to exploit existing high-quality samples more effectively. To improve GRPO training, samples are then organized into an error-aware curriculum, with a larger proportion of medium and hard examples, enabling the model to focus more on challenging cases while still maintaining a gradual learning progression rather than being randomly subsampled. Finally, inspired by thought dropout~\cite{wang2025think}, we introduce a guided thought dropout mechanism that selectively removes CoT for cases already solved by the pretrained model, thereby making more effective use of CoT annotations. This design offers a more principled alternative to random dropout. 
Experiments demonstrate that Omni-CLST achieves a competitive 73.80\% on the publicly available MMAU-mini benchmark and establishes a new state of the art with 64.30\% on MMAR, validating the effectiveness of error-aware curriculum learning and guided selective CoT.

In summary, our work makes three main contributions:
\begin{itemize}
\vspace{-1mm}
\item We introduce an error-aware curriculum learning para-digm that focuses on challenging, error-prone samples, enhancing the effectiveness of curriculum-based training for AQA.
\vspace{-1mm}
\item We propose a guided thought dropout mechanism that selectively skips the CoT process for easier samples while retaining reasoning for harder cases, enabling more efficient utilization of datasets containing explicit reasoning traces.
\vspace{-1mm}
\item We achieve a competitive 73.80\% on MMAU-mini and establishes a new state of the art with 64.30\% on MMAR, validating the effectiveness of our approach. Code and models are available at \url{https://github.com/NKU-HLT/Omni-CLST}.
\end{itemize}

\begin{figure*}[t] 
\centering
\includegraphics[width=\textwidth]{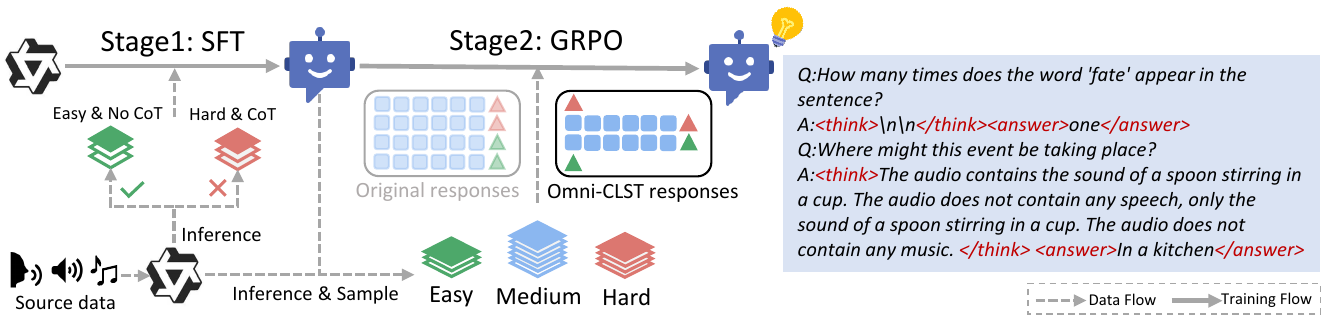}
\caption{The overall framework of Omni-CLST. In the Original responses and Omni-CLST response,
\includegraphics[height=1em]{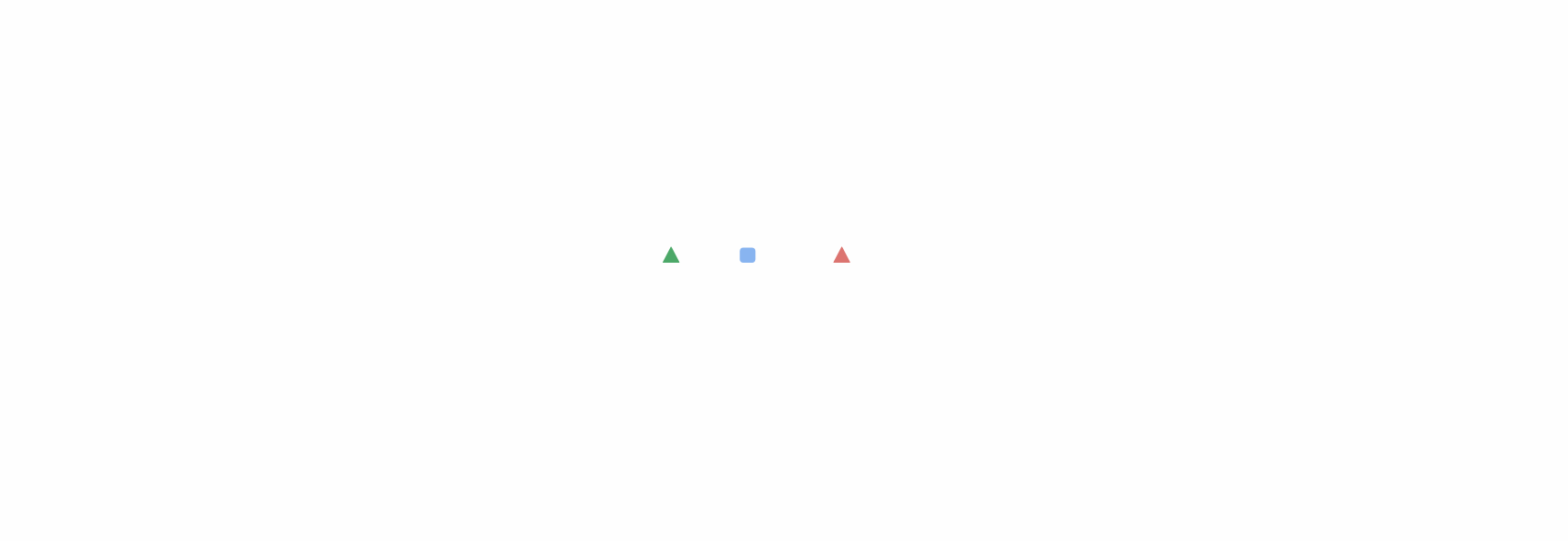} denote CoT tokens,
while \includegraphics[height=1em]{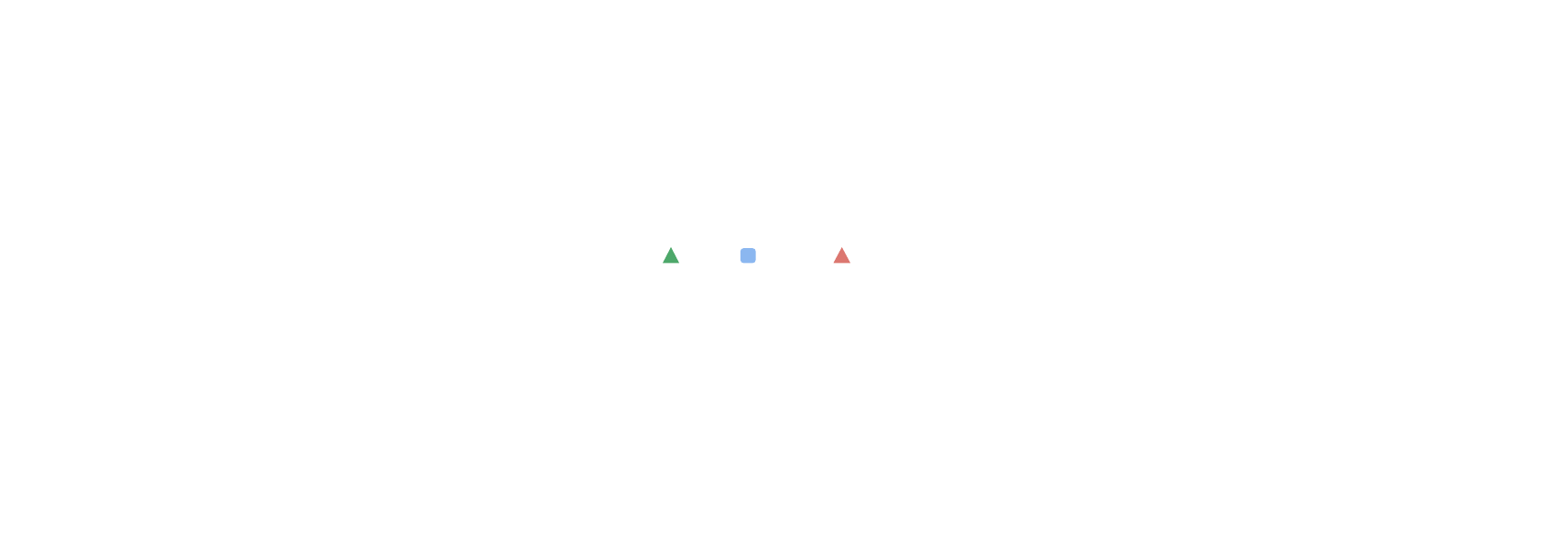} and
\includegraphics[height=1em]{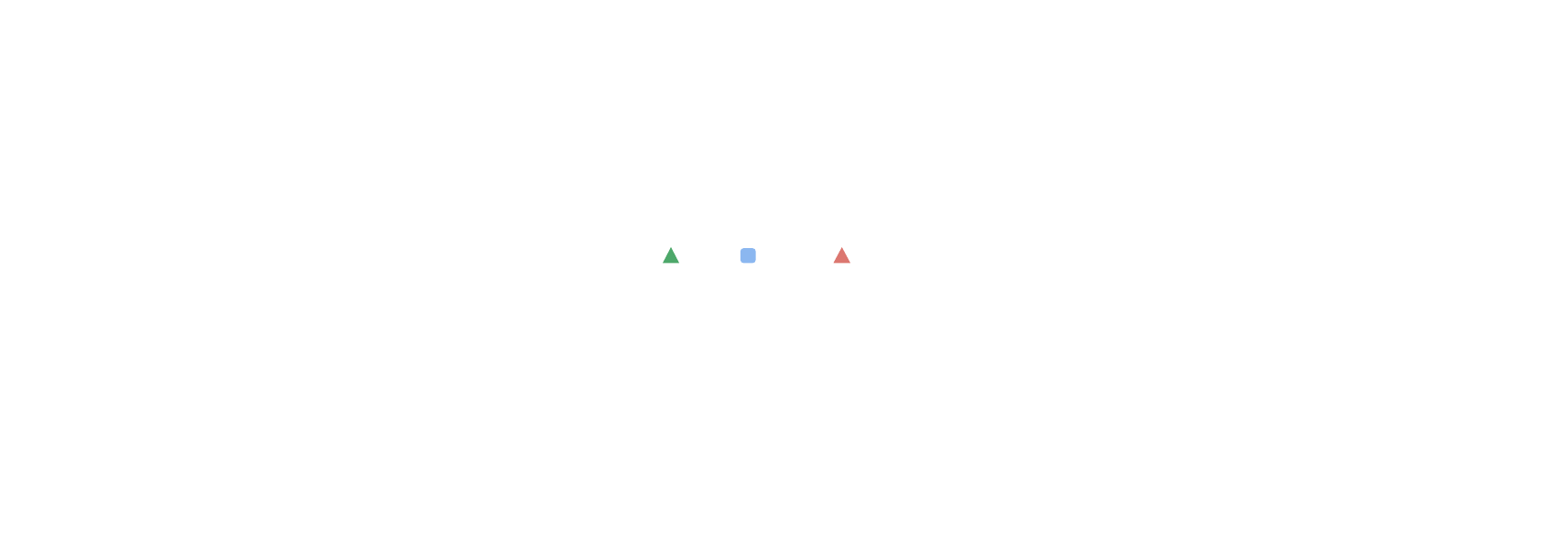} represent incorrect and correct answers, respectively.
}
\label{fig:model}
\vspace{-3mm}
\end{figure*}

\section{Method}
\label{sec:pagestyle}
\subsection{Overall Framework}

Figure~\ref{fig:model} illustrates the overall framework of Omni-CLST, which consists of two stages: SFT and GRPO. In the SFT stage, the pretrained Qwen2.5-Omni model first performs inference on the training set, dropping CoT for correctly answered samples while retaining it for those answered incorrectly. Based on the outcomes of the pretrained and SFT models, samples are organized into three difficulty levels: easy, medium, and hard. This structured curriculum allows the model to progressively focus on more challenging cases. In the subsequent GRPO stage, the model learns to autonomously determine when to engage in CoT, further enhancing its reasoning capabilities on AQA tasks.

\subsection{Error-aware Curriculum Learning Paradigm}
To better exploit the training data, we design an error-aware curriculum learning paradigm that organizes samples into progressive difficulty levels. Specifically, we categorize data according to the responses of the pretrained model and the SFT model. Samples correctly answered by the pretrained model are considered easy, those corrected by the SFT model after being initially answered incorrectly are regarded as medium, and those still answered incorrectly even after SFT are treated as hard. We found that medium and hard examples are more valuable for GRPO training. Therefore, they constitute a larger proportion of the training data, enabling the model to focus on more informative and challenging cases while still maintaining a progression from easy to complex.

\subsection{Guided Thought Dropout Mechanism}
In the SFT stage, we guide the model to learn when to perform reasoning. 
Specifically, if the pretrained model answers a question correctly, we drop its CoT; 
if it answers incorrectly, we retain the CoT. 
In this way, the model is explicitly guided during SFT to associate reasoning with harder cases.

During the GRPO stage, given an audio $a \in \mathcal{A}$ and a text query $q \in \mathcal{Q}$, we sample $N$ candidate responses $\{o_1,o_2,\ldots,o_N\}$ from the current policy $\pi_{old}$. Each response $o_i$ is evaluated with a reward function $r(\cdot)$, and the rewards are normalized by mean and standard deviation to define the advantage:  

\vspace{-4mm}
\begin{equation}
A_i = \frac{r(o_i) - \mathrm{mean}\{r(o_1), r(o_2), \ldots, r(o_N)\}}{\mathrm{std}\{r(o_1), r(o_2), \ldots, r(o_N)\}}
\end{equation}
where $A_i$ represents the relative advantage of candidate response $o_i$ compared with other responses in the group.  

The policy $\pi_\theta$ is then updated with the following objective, which encourages the model to generate responses with higher within-group advantages:  

\vspace{-6mm}
\begin{equation}
\begin{aligned}
J_{\mathrm{GRPO}}(\theta) 
&= \mathbb{E}_{\{o_i\}_{i=1}^N \sim \pi_{old}(a,q)} \frac{1}{N} \sum_{i=1}^N \Big( 
   \min\!\left[\alpha_i \cdot A_i,\right. \\
& \quad \left. \mathrm{clip}\!\left(\alpha_i, \, 1-\epsilon, \, 1+\epsilon\right) \cdot A_i \right] 
   - \beta \, \mathbb{D}_{\mathbb{KL}}\!\big[\pi_\theta \,\|\, \pi_{ref}\big] \Big)
\end{aligned}
\label{eq:grpo_obj}
\end{equation}
where $\alpha_i = \frac{\pi_\theta(o_i|a,q)}{\pi_{old}(o_i|a,q)}$. In Eq.~\ref{eq:grpo_obj}, the guided thought dropout, applied during the GRPO stage, 
can reshape the latent distribution of responses $\pi_\theta(o_i|a,q)$, 
thereby enhancing the diversity of $\alpha_i$. 
Compared to the ``random'' thought dropout strategy used in the SFT stage, 
guided thought dropout promotes a better reasoning trajectory.

We employ two reward functions during GRPO training. 
The first is the accuracy reward $R_{\text{acc}}$, which assigns $1$ if the predicted answer $o_i$ matches the ground truth $s_i$, and $0$ otherwise:

\vspace{-2mm}
\begin{equation}
R_{\text{acc}}(o_i, s_i) = 
\begin{cases}
1, & \text{if } o_i = s_i, \\
0, & \text{otherwise},
\end{cases}
\label{eq:acc_reward}
\end{equation}
The second is the format reward $R_{\text{f}}$, which verifies whether the output strictly follows the required format $\mathcal{F}$. $\mathcal{F} = \texttt{<think>...</think><answer>...</answer>}$
\begin{equation}
R_{\text{f}}(o_i) = 
\begin{cases}
1, & \text{if } o_i \in \mathcal{F}, \\
0, & \text{otherwise}.
\end{cases}
\label{eq:fmt_reward}
\end{equation}
The final reward is a weighted combination of these two signals:
\begin{equation}
R_i = \lambda_{\text{acc}} \, R_{\text{acc}}(o_i, s_i) 
    + \lambda_{\text{f}} \, R_{\text{f}}(o_i),
\label{eq:final_reward}
\end{equation}
where $\lambda_{\text{acc}}$ and $\lambda_{\text{f}}$ are hyperparameters controlling the contribution of accuracy and format rewards, respectively.

\section{Experiment and results}
\label{sec:typestyle}

\begin{table*}[t]
\centering
\caption{Performance of open-source AQA models on MMAU-mini (v05.15.25) and MMAR. Models marked with $\dagger$ denote results obtained from the official MMAU-mini and MMAR benchmark reproductions. Omni-CL-CoT (Ours) refers to our model trained with only the error-aware curriculum learning paradigm, where all questions use the CoT process. All results are reported in terms of accuracy (\%). For MMAR, Mix1–Mix4 correspond to different modality combinations: sound–music, sound–speech, music–speech, and sound–music–speech, respectively.}
\label{tab:result}
\vspace{0.5mm}
\resizebox{\textwidth}{!}{
\begin{tabular}{l|cccc|cccccccc}
\toprule
\multirow{2}{*}{\textbf{Model}} 
& \multicolumn{4}{c|}{\textbf{MMAU-mini}} 
& \multicolumn{8}{c}{\textbf{MMAR}} \\
\cmidrule(lr){2-5} \cmidrule(lr){6-13}
& \textbf{Sound} & \textbf{Music} & \textbf{Speech} & \textbf{Avg} 
& \textbf{Sound} & \textbf{Music} & \textbf{Speech} & \textbf{Mix1} 
& \textbf{Mix2} & \textbf{Mix3} & \textbf{Mix4} & \textbf{Avg} \\
\midrule
Qwen2-Audio 7B$\dagger$\cite{chu2024qwen2} & 67.27  & 56.29  & 55.26  & 59.60 
&  33.33 & 24.27 & 32.31 & 9.09 & 31.19 & 30.49 & 25.00 & 30.00 \\
Audio Reasoner$\dagger$\cite{xie2025audio} & 67.87  & 69.16 & 66.07 & 67.70
& 43.64 &  33.50 & 32.99 & 45.45 & 42.66 &  31.71 &  25.00 & 36.80 \\
R1-AQA\cite{li2025reinforcement} & 75.68 & 66.47  & 63.66  & 68.60 
& 58.18 & 39.81 & 58.18 & \textbf{63.64} & 60.55 & 53.66 & 50.00 & 52.60 \\
Qwen2.5-Omni 7B$\dagger$\cite{xu2025qwen2} & \underline{78.10}	 & 65.90 & 70.60 & 71.50  
& 58.79 &  40.78 & 59.86 & \underline{54.55} &  61.93 & \textbf{67.07} & 58.33 & 56.70 \\
AudSemThinker\cite{wijngaard2025audsemthinker} & 77.18  & 64.67 & \underline{68.17} & 70.00 
& 52.73 & 41.26 & 52.72 & 27.27 & 56.42 & 56.10 & 37.50 & 50.80 \\
Audio-Flamingo 3$\dagger$\cite{goel2025audio} & \textbf{79.58} & 66.77 & 66.37  & \underline{73.30}
& 53.94 & 49.51 & 62.59 & 36.36 & 65.60 & 62.20 & 54.17 & 58.60 \\
Omni-CL-CoT (Ours) & 78.08  & \textbf{74.25} & 67.27 & 73.20 & \underline{64.24} & \underline{51.94} & \textbf{66.67} & \underline{54.55} & \underline{68.35} & \underline{64.63} & \underline{62.50} & \underline{63.20} \\
\textbf{Omni-CLST (Ours)} & 75.98 & \underline{73.35} & \textbf{72.07}& \textbf{73.80}  & \textbf{67.27} & \textbf{53.88} & \underline{64.29} & \textbf{63.64} & \textbf{72.02} & 62.20 & \textbf{70.83} & \textbf{64.30} \\
\bottomrule
\end{tabular}}
\vspace{-3mm}
\end{table*}

\subsection{Datasets}

We conduct our experiments using two high-quality audio-based datasets, which provide rich caption or CoT information. The first dataset is derived from \textbf{AVQA}~\cite{10.1145/3503161.3548291}, a manually designed benchmark for audio-visual question answering in realistic scenarios. We retain only the audio-text pairs and replace occurrences to ``video'' in the questions with ``audio''. The resulting AVQA training set contains approximately 36k samples, with a validation set of about 15k samples. We utilize the captions provided by \textbf{AudioSetCaps}\cite{bai2025audiosetcaps} for each audio clip as the CoT content. The second dataset is \textbf{AudSemThinker}~\cite{wijngaard2025audsemthinker}, which leverages outputs from five pretrained models to analyze crawled YouTube audio and aggregates high-quality CoT. This multi-choice QA dataset contains around 213k training samples.

\subsection{Experimental Setup}
During the SFT stage, we fine-tune the model with LoRA~\cite{hu2022lora}, using a learning rate of $1\mathrm{e}{-4}$, LoRA rank of $8$, LoRA $\alpha$ of $32$, for $2$ epochs with a batch size of $6$. In the GRPO stage, we sample training data according to model performance, with a ratio of pretrained model correct : SFT correct : SFT incorrect = 3:20:7, using 15k samples for one epoch. We train with a batch size of $1$, a temperature of $1.2$, $4$ candidate responses per GRPO step, a learning rate of $1\mathrm{e}{-6}$, and $\beta = 0.04$ in Equation~\ref{eq:grpo_obj}. $\lambda_{\text{acc}} = 2$ and $\lambda_{\text{f}} = 1$ in Equation~\ref{eq:final_reward}. All experiments are conducted on $4$ NVIDIA A100 GPUs (80GB). 

\subsection{Benchmarks}
We evaluate model on the following two AQA benchmarks.

\textbf{MMAU}~\cite{DBLP:conf/iclr/SakshiTKSSNDGM25} is a benchmark for multimodal audio understanding, containing 10k carefully curated audio clips paired with human-annotated questions and answers across speech, environmental sounds, and music. For analysis, MMAU is further split into a smaller subset (\textbf{MMAU-mini}) with 1,000 samples.

\textbf{MMAR}~\cite{ma2025mmar} is designed to assess the deep reasoning capabilities of audio-language models across a wide range of real-world audio scenarios. It consists of 1,000 high-quality audio-question-answer triplets, requiring multi-step reasoning and domain-specific knowledge.

\subsection{Main Results and Analysis}

Table~\ref{tab:result} summarizes the performance of open-source AQA models on MMAU-mini and MMAR benchmarks. Overall, our proposed approaches, Omni-CL-CoT and Omni-CLST, achieve competitive or superior results compared with existing open-source models.

Omni-CLST achieves an average accuracy of 73.8\% on MMAU-mini, while on the more challenging MMAR benchmark it reaches 64.3\%. This performance is attained through our error-aware curriculum learning paradigm combined with guided selective CoT, without the need for additional QA construction pipelines. For reference, the recently proposed Omni-R1 reports 77.0\% on MMAU-mini and 63.4\% on MMAR, but relies on ChatGPT-generated QA data from AVQA/VGG-Sound~\cite{Chen20} captions.

Further analysis provides additional insights into the effectiveness of our framework. First, we find that 67.6\% of MMAR samples and 68.7\% of MMAU-mini samples can be solved without invoking the CoT process. By adaptively deciding when to reason step-by-step, Omni-CLST achieves higher accuracy while avoiding unnecessary reasoning overhead. Second, we observe that Omni-CL-CoT produces an average of 84.25 tokens on MMAU-mini and 70.14 tokens on MMAR, whereas Omni-CLST reduces the averages to 38.12 and 30.48 tokens, respectively, demonstrating the efficiency and superiority of the guided thought dropout strategy.

A concrete example is counting, a task known to be challenging for LLMs. In MMAR, among 99 counting and statistic questions, Omni-CLST improves the accuracy from 43.43\% (Omni-CL-CoT) to 48.48\%. Notably, this gain primarily comes from cases where the model correctly answers without invoking the reasoning process, suggesting that the CoT step can sometimes interfere with questions the model could solve directly.

\subsection{Ablation Studies}

We conduct ablation experiments to evaluate the key components of our framework, as summarized in Table~\ref{tab:ablation}.

(A) examines the effectiveness of the SFT phase and the CoT process under randomly sampled training data. We observe that introducing the CoT process improves performance on both benchmarks regardless of whether SFT is applied. Moreover, for the more challenging MMAR benchmark, performing SFT prior to GRPO yields better results than directly applying GRPO, indicating the benefit of the preliminary SFT stage for subsequent policy optimization.

(B) assesses the effectiveness of our error-aware curriculum learning paradigm. By organizing GRPO training samples according to difficulty, with a larger proportion of medium-difficulty samples and a suitable fraction of hard examples, we find that the model achieves improved learning efficiency and higher accuracy, demonstrating the value of curriculum-based data organization. 

(C) investigates the thought dropout strategy. Under random thought dropout, the model achieves results comparable to using CoT for all samples (i.e., Omni-CL-CoT). However, when guided thought dropout is applied, the performance improves significantly, confirming the effectiveness of our guided dropout strategy in selectively leveraging CoT supervision.

\begin{table}[t]
\centering
\caption{Ablation study. (A) Effectiveness of SFT and CoT uses 15k randomly sampled training data. 
``Pretrained-based" refers to GRPO training data sampled with a ratio of pretrained model correct : incorrect = 1:5. Results are reported as average accuracy (\%).}
\label{tab:ablation}
\vspace{1mm}
\resizebox{\linewidth}{!}{
\begin{tabular}{l|c|c}
\toprule
\textbf{Setup} & \textbf{MMAU-mini} & \textbf{MMAR} \\
\midrule
\multicolumn{3}{c}{\textit{(A) Effectiveness of SFT and CoT}} \\
\midrule
GRPO & 72.5 & 56.9 \\
GRPO + CoT & \textbf{75.1} & 59.2 \\
SFT + GRPO & 72.8 & \underline{61.3} \\
SFT + GRPO + CoT & \underline{73.6} & \textbf{62.1} \\
\midrule
\multicolumn{3}{c}{\textit{(B) GRPO Training Data Curriculum}} \\
\midrule
Pretrained-based & 71.4 & \textbf{63.4} \\
Omni-CL-CoT (Ours) & \textbf{73.2} & 63.2 \\
\midrule
\multicolumn{3}{c}{\textit{(C) Thought Dropout Strategy}} \\
\midrule
Omni-CL + random thought dropout & 73.1 & 63.5 \\
Omni-CLST (Ours) & \textbf{73.8} & \textbf{64.3} \\
\bottomrule
\end{tabular}}
\end{table}
\section{CONCLUSION}
\label{sec:CONCLUSION}

We propose Omni-CLST, an error-aware curriculum learning framework with guided selective Chain-of-Thought reasoning for audio question answering. The model efficiently leverages high-quality training data and focuses CoT on challenging cases. Experiments demonstrate that Omni-CLST achieves a competitive 73.80\% on the publicly available MMAU-mini benchmark and establishes a new state of the art with 64.30\% on MMAR, validating the effectiveness of error-aware curriculum strategies and selective CoT in enhancing multimodal audio-language models. In summary, Omni-CLST provides an effective method to maximize the utility of high-quality datasets through error-aware curriculum and guided selective CoT, enabling more efficient exploitation of informative reasoning signals.

{
  \small
\bibliographystyle{IEEEbib}
\bibliography{strings,refs}
}

\end{document}